\documentclass[aps,superscriptaddress,twocolumn,twoside,floatfix,pra,nofootinbib,a4paper]{revtex4}
\usepackage{graphicx,amsmath,amssymb,color}
\usepackage[normalem]{ulem}
\usepackage{amsfonts}
\usepackage[ocgcolorlinks,colorlinks=true,linkcolor=blue,citecolor=red]{hyperref}
\usepackage{latexsym}
\usepackage{amsfonts}
\usepackage{mathrsfs}
\usepackage{natbib}
\usepackage{color,verbatim}
\DeclareMathAlphabet{\mathrsfs}{U}{rsfs}{m}{n}
\DeclareMathAlphabet{\mathpzc}{OT1}{pzc}{m}{it}
\DeclareMathAlphabet{\matheus}{U}{eus}{m}{n}
\DeclareMathAlphabet{\mathbbold}{U}{bbold}{m}{n}

\newcommand{\ba}{\begin{eqnarray}}
\newcommand{\be}{\begin{equation}}
\newcommand{\ee}{\end{equation}}

\newcommand{\ea}{\end{eqnarray}}
\newcommand{\ban}{\begin{eqnarray*}}
	\newcommand{\ean}{\end{eqnarray*}}

\begin{document}
	
	\title{Bell-type Inequalities for Arbitrary Non-Cyclic Networks}
	
	\author{Armin Tavakoli}
	\affiliation{Department of Physics, Stockholm University, S-10691 Stockholm, Sweden}
	\affiliation{Computer Science Division, University of California, Berkeley, California 94720, USA}
	\affiliation{Department f\"ur Physik, Ludwig-Maximilians-Universit\"at, D-80797 M\"unchen, Germany}

	
	\date{\today}

	
	\begin{abstract}
		Bell inequalities bound the strength of classical correlations between observers measuring on a shared physical system. However, studies of physical correlations can be considered beyond the standard Bell scenario by networks of observers sharing some configuration of many independent physical systems. Here, we show how to construct Bell-type inequalities for correlations arising in any tree-structured network i.e. networks without cycles. This is achieved by an iteration procedure that in each step allows one to add a branch to the tree-structured network and construct a corresponding Bell-type inequality. We explore our inequalities in several examples, in all of which we demonstrate strong violations from quantum theory.   
	\end{abstract}
	
	
	
	\maketitle
	\textit{Introduction.---}
	The milstone work of John Bell showed that quantum correlations  arising between spatially separated observers can break the limitations of classical physics \cite{Bell64}. Studies of correlations predicted by quantum theory has been a key to understanding fundamental properties of the theory and has led to applications in information processing including random number generation \cite{PA10}, device-independent cryptography \cite{DI} and reduction of communication complexity \cite{BC10}.  
	
\begin{figure}
	\centering
	\includegraphics[width=\columnwidth]{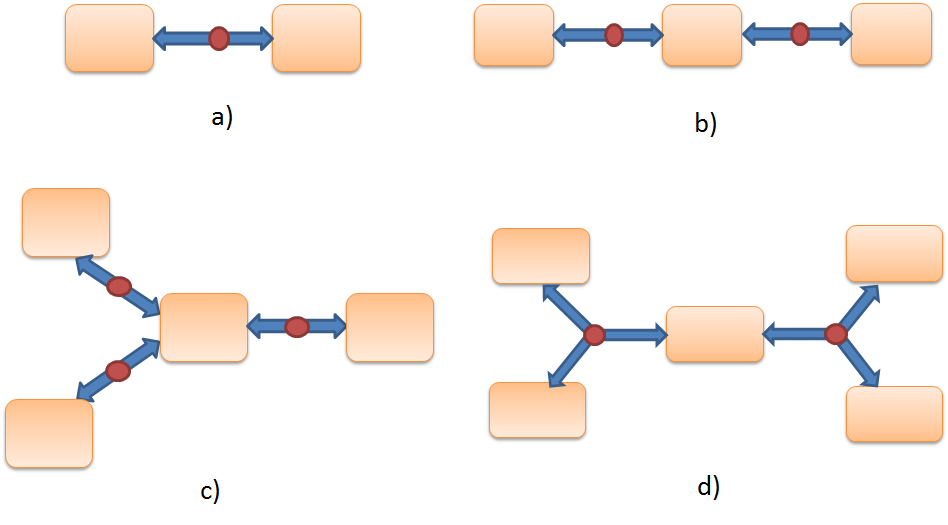}
	\caption{a) Standard Bell experiment. b) Chain-network. c) Star-network. d) Multipartite star-network}
	\label{fig1}
	\end{figure}

	A Bell experiment considers a source emitting a physical system shared between a set of observers who can randomly chose to apply some local measurements. The standard two-particle Bell experiment is illustrated in Figure \ref{fig1} a). The properties of the correlations arising between the outcomes of the observers in Bell experiments have been thoroughly studied \cite{Bell14} and may be considered fairly well understood. However, much less is known about the nature of correlations arising in more sophisticated network configurations beyond the Bell experiments. A network could in general involve many independent sources each emitting a physical system that is shared between some set of observers. Importantly, a single observer could be receiving subsystems of many independent physical systems originating from different sources. 
	
	There are both conceptual and applied motivations for studying classical and quantum correlations in networks. Networks naturally generalize Bell experiments which makes them conceptually attractive. Also, the notion of classical correlations on a network leads to stronger constraints than those associated to standard Bell inequalities. However, these constraints also influence the strength of quantum correlations, which makes it interesting to study their comparative strength as opposed to standard Bell inequalities. 
	Furthermore, networks are relevant in a variety of applications involving entanglement swapping experiments \cite{ZZHE93}, entanglement percolation \cite{ACL07} and quantum repeaters protocols \cite{SSRG11, SB05}. Quantum correlations in networks are interesting for the practical implementation of large scale quantum communication networks which is arguably one of the main goals of applied quantum information.

	Bell's theorem can be viewed as a statement of causal inference: by observing correlations we can infer a conclusion on the nature of the cause generating the correlations. The approach of causal inference has been extensively used to analyze correlations in networks \cite{F12(1), F12(2), WS12, HLP14, CK15, LS15}. Furthermore, Bell-type inequalities have been derived for various classes of networks.
	The first such analysis considered a chain-network of three observers involving two two-particle physical sources \cite{BGP10, BRGP12}, see Figure \ref{fig1} b). This was generalized by studies of Bell-type inequalities for networks in a star-shape configuration \cite{TSCA14}, an example of which is given in Figure \ref{fig1} c). The Bell-type inequalities for star-networks have been extended to networks with sources emitting multipartite physical systems \cite{T15}, see e.g. Figure \ref{fig1} d). Methods for deriving network Bell-type inequalities have been more broadly analyzed in Ref.\cite{C15}. Furthermore, an iterative approach has been presented for finding Bell-type inequalities on networks \cite{RB15}: given a Bell-type inequality for some network, one can construct a new Bell-type inequality for the same network to which we have now added a source connecting some observer in the initial network to one new observer. This method allows for finding Bell-type inequalities on cycle-free networks that involve only two-particle physical systems, up to variations of the initial network and the Bell inequality from which the iteration initiates. 
	
	In this work, we will show how to construct Bell-type inequalities for any tree-structured network i.e. any network that does not contain cycles. This will be achieved by an iterative procedure in which the network and the associated Bell-type inequality are gradually constructed. Explicitly, we will show that if given a Bell-type inequality for some network $\mathscr{N}$, we can construct a new Bell-type inequality for another network $\mathscr{N}'$ which is obtained by adding to $\mathscr{N}$ a new source connecting one observer to $L$ new observers. Importantly, we will give many examples of networks in which quantum theory violates our inequalities. We will both reproduce the strength of  quantum violations on some already studied networks, and demonstrate strong quantum violations on networks for which our method makes a contribution.

\textit{Network Bell-type inequalities.---}	
	A network $\mathscr{N}$ consists of $N$ sources, the $j$'th of which is associated to a local random variable $\lambda_j$, belonging to some space $\Lambda_j$, with some density function $\rho_j(\lambda_j)$. Let there be $M$ observers $A^1,\ldots,A^M$ in $\mathscr{N}$. The $k$'th observer can chose a measurement labeled by $X_k$ from a set of choices and the outcome is denoted $a^k\in\{0,1\}$. In a classical model of the resulting probability distribution $P(a^1,\ldots,a^M|X_1,\ldots,X_M)$, the local outcome of $A^k$ is determined by $X_k$ and the set of local random variables $\bar{\lambda}_k$ to the sources of which $A^k$ is connected by receiving a part of the emitted physical system. Thus, any classical model satisfies \cite{RB15};
	\begin{multline}\label{classical}
		P(a^1,...,a^M|X_1,...,X_M)=\\
		\int_{\Lambda_1}d\lambda_1 \rho_1(\lambda_1)\ldots \int_{\Lambda_N}\!\!\!d\lambda_N\rho_N(\lambda_N)\prod_{k=1}^{M} P(a^k|X_k,\bar{\lambda}_k).
	\end{multline}
	Note that this definition naturally extends the previously constructed definitions of classical correlations on networks \cite{BGP10, BRGP12, TSCA14}.

	Let us now show how to construct Bell-type inequalities on any network that is a tree-structure; that is arbitrary cycle-free networks. Assume that on $\mathscr{N}$ we have a Bell inequality, which can generally be written on the form 		
	\begin{equation}\label{Bell}
	\sum_{X_1\ldots X_M} c_{X_1\ldots X_M}\langle a^1_{X_1}\ldots a^M_{X_M}\rangle \leq 1
	\end{equation} 
	where $a^k_{X_k}$ denotes the $X_k$'th measurement of $A^k$, $c_{X_1...X_M}$ are some real coefficients and $\langle a^1_{X_1}\ldots a^M_{X_M}\rangle $ denotes the global correlator defined as $\langle a^1_{X_1}\ldots a^M_{X_M}\rangle=\!\!\!\sum_{a^1...a^M}(-1)^{\sum_{j=1}^{M}a^j}P(a^1,...,a^M|X_1,...,X_M)$.

	We construct a new network $\mathscr{N}'$ in which we have extended $\mathscr{N}$ by connecting to observer $A^M$ a source which emits a physical system shared between $A^M$ and $L$ new observers $B^1,...,B^L$ among whom the $l$'th observer performs one of two measurements, either $b^l_0$ or $b^l_1$, for $l=1,\ldots, L$. Let $A^M$ chose between $2^{s_M}$ measurements, for some $s_M\geq 0$. We aim to find a Bell-type inequality on $\mathscr{N}'$. For simplicity, we divide the possible scenarios into two:
	
	\textbf{Case 1:} $s_M \geq L$. We will group $A^M$'s measurements into $2^L$ disjoint non-empty sets $\kappa_X$ where the index $X$ runs over all subsets $X\subset \mathbb{N}_L\equiv \{1,...,L\}$. Then, for some set of positive real numbers $\{q_X\}_{X\subset \mathbb{N}_L}$ satisfying $\sum q_X=1$, there exists a Bell-type inequality for correlations on $\mathscr{N}'$:
	\begin{equation}\label{ineq}
	\sum_{X\subset \mathbb{N}_L} \frac{Q_X}{q_X}\leq 1 
	\end{equation}
	where we have defined
	\begin{equation}\label{QX}
	Q_X=\!\!\!\!\!\!\sum_{\substack{X_1...X_{M-1}\\X_M \in \kappa_{X}}}\!\!\!\!\!\!\! c_{X_1...X_M} \left\langle a^1_{X_1}...a^M_{X_M}\prod_{k=1}^{L}\frac{b^k_0+(-1)^{\delta^k_X}b^k_1}{2}\right\rangle \leq 1
	\end{equation}
	where $\delta^k_X=1$ if $k\in X$ and otherwise $\delta^k_X=0$. 
	
	To prove this claim, we associate a hidden variable $\lambda_j\in \Lambda_j$ with density function $\rho_j(\lambda_j)$ to the $j$'th source in $\mathscr{N}$. The source we add to $\mathscr{N}$ is associated to hidden variable $\mu$, belonging to some space $W$, with density function $\rho_W(\mu)$. Since the $L$ new observers have binary outcomes, they have together $2^L$ possible outcome strings. For every possible outcome string, the associated values of $\mu$ are collected in the sets $W_X \equiv \{\mu\in W|b_0^k=(-1)^{\delta^k_X}b^k_1, k=1,...,L\}$. Clearly $\{W_X\}_X$ is a partition of $W$. The probability associated to $\mu\in W_X$ is denoted $q_X=\int_{W_X} d\mu\rho_W(\mu)$ from which it holds that $\sum q_X=1$. Furthermore, we introduce  distributions $\rho_{W_X}=\rho_W/q_X$ normalized on $W_X$. 
	
	In a classical model, $Q_X$ is written:
	\begin{multline}\label{Q}
	Q_X=\sum_{\substack{X_1...X_{M-1}\\ X_M\in \kappa_X}}\!\!\!\!\!\!\!\!c_{X_1...X_M}\left\langle a^1_{X_1}...a^M_{X_M}  \prod_{k=1}^{L} \frac{b^k_0+(-1)^{\delta^k_X}b^k_1}{2}\right\rangle\\
	=\!\!\!\!\!\!\sum_{\substack{X_1...X_{M-1}\\ X_M\in \kappa_X}}\!\!\!\!\!\!c_{X_1...X_M}\int_{\Lambda_1}d\lambda_1\rho_1(\lambda_1)\ldots\!\! \int_{\Lambda_N}d\lambda_N\rho_N(\lambda_N) \int_{W}d\mu \rho_W(\mu)\\
	\times \prod_{l=1}^{M-1}a^l_{X_l}(\bar{\lambda}_l)\times a^M_{X_M}(\bar{\lambda}_M,\mu) 
	 \prod_{k=1}^{L} \frac{b^k_0(\mu)+(-1)^{\delta^k_X}b^k_1(\mu)}{2}\\
	=q_X \!\!\! \!\!\!\sum_{\substack{X_1...X_{M-1}\\ X_M\in \kappa_X}}\!\!\!\!\!\!c_{X_1...X_M}
	 \int_{W_X} d\mu_{X} \rho_{W_X}(\mu_X) \langle a^1_{X_1}...a^{M-1}_{X_{M-1}}\hat{a}^M_{X_M,X,\mu_X}\rangle 
	\end{multline} 
	where $\hat{a}^M_{X_M,X,\mu_X}=a^M_{X_M}(\bar{\lambda}_M,\mu_X)\prod_{k=1}^{L}b^k_0$ if $\mu_X\in W_X$ and otherwise $\hat{a}^M_{X_M,X,\mu_X}=0$. Note that we have relabeled the integration variable $\mu$ as $\mu_X$ only for sake of clarity. We can think of  the product series over $b^k_0$ as a trivial relabeling of the correlator $\langle a^1_{X_1}...a^{M}_{X_{M}}\rangle $ on $\mathscr{N}$ by either positive or negative sign. 
	
	Define quantities
	\begin{equation}
	S_{X,\mu_X}\equiv \sum_{\substack{X_1...X_{M-1}\\ X_M\in \kappa_X}}c_{X_1...X_M}\langle a^1_{X_1}...a^{M-1}_{X_{M-1}}\hat{a}^M_{X_M,X,\mu_X}\rangle.
	\end{equation}
	Note that equation \eqref{Q} now can be written 
	\begin{equation}\label{Qf}
	Q_X=q_X\int_{W_X}d\mu_X \rho_{W_X}(\mu_X)S_{X,\mu_X}.
	\end{equation}

	Due to our understanding of the correlators in $S_{X,\mu_X}$ as relabelings of the original correlators on $\mathscr{N}$, it follows directly from \eqref{Bell} that
	\begin{equation}\label{sumS}
	\sum_{X\subset \mathbb{N}_L}S_{X,\mu_X}\leq 1.
	\end{equation}
	Dividing both sides of \eqref{Qf} with $q_X$ and summing both sides of \eqref{Qf} over all $X\subset \mathbb{N}_L$ and using \eqref{sumS} together with the fact that $\rho_{W_X}$ is a normalized distribution on $W_X$, we obtain the expression \eqref{ineq}. Thus, there exists a set $\{q_X\}_X$  such that \eqref{ineq} is a Bell-type inequality on $\mathscr{N}'$.
	
	\textbf{Case 2:}  $s_M<L$. Since we can no longer partition the set of $A^M$'s measurements into $2^L$ non-empty sets, we will increase the number of measurements of $A^M$ so that $s_M=L$. Again, indexing the measurements of $A^M$ by the subsets of $\mathbb{N}_{L}$, we can make the trivial partition $\kappa_X=\{X\}$ for every $X\subset \mathbb{N}_L$. To find a Bell-type inequality, we again follow the above procedure, with the minor modification that the right-hand side of \eqref{sumS} is replaced with $2^{L-s_M}$ since we have made use of the inequality \eqref{Bell} $2^{L-s_M}$ times. Thus we find the inequality:
	\begin{equation}\label{ineq'}
	\sum_{X\subset \mathbb{N}_L} \frac{Q_X}{q_X}\leq 2^{L-s_M}.
	\end{equation}
	
	

	A first remark: above we assumed that all $q_X$ are positive real numbers whereas nothing in the above arguement prevents some $q_X$ to be zero, yielding singularities in \eqref{ineq} and \eqref{ineq'}. Therefore, it is necessary to note that in such cases the inequalities \eqref{ineq} and \eqref{ineq'} will be modified such that the summation in the left-hand-side only goes over events with $q_X>0$ which are physically relevent. 
	
	A second remark: the restriction to $A^M$ having $2^{s_M}$ measurements is not a necessary condition, but is introduced only for simplicity. If $A^M$ were to have $s_M'$ measurements instead of $2^{s_M}$, then there is modification of the above only when $s_M'<2^L$: increase the number of measurement options from $s_M'$ to $LCM(s_M',2^L)$, where $LCM$ denotes the least common multiple, and use case 1 repeatingly as described in case 2 above. The bound in \eqref{ineq'} changes to $LCM(s_M',2^L)/s_M'$. Note that whenever $s_M'=2^{s_M}$ and $s_M\leq L$, then $LCM(2^{s_M},2^L)/2^{s_M}=2^{L-s_M}$ which is the bound in \eqref{ineq'}.  
	
	A third remark: With minor modification following \cite{RB15}, we can involve non-full correlation terms in our inequalities. Also, it is worth observing that when $\mathscr{N}'$ is an extension of $\mathscr{N}$ by a two-particle source ($L=1$), our inequalities reduce to those derived in Ref.\cite{RB15}, up to the just mentioned minor modifications.  
	
	 However, the relevance of our Bell-type inequalities ought to be judged  from the possibility of violating them by a theory in which probability distributions do not admit the form \eqref{classical}. Of particular interest are the physically realizable distributions, obtained from quantum theory. Therefore, we will proceed with several examples of networks for which we explicitly find Bell-type inequalities and demonstrate violations by quantum thery. To gain confidence in the relevance of our inequalities, we will both use our method to reproduce results previously known, and also go beyond the scope of previous work.

\begin{figure}
		 		\centering
		 		\includegraphics[width=\columnwidth]{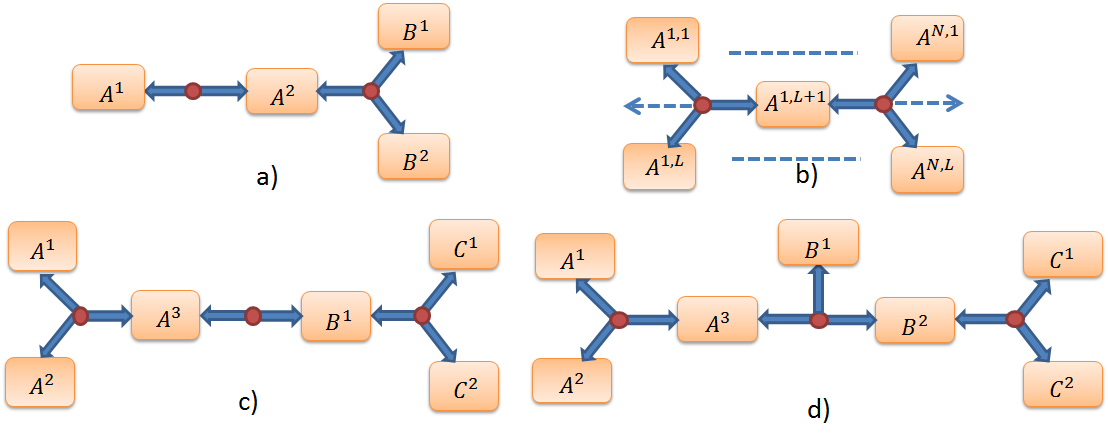}
		 		\caption{The networks in examples 1, 2, 3 and 4.}
		 		\label{Examples}
		 	\end{figure}

	\textit{Example 1.---}
	We consider the network $\mathscr{N}'$ in Figure \ref{Examples} a). We construct $\mathscr{N}'$ by starting from the network $\mathscr{N}$, involving only $A^1$ and $A^2$, and then connect $A^2$ via a three-particle source to the new observers $B^1$ and $B^2$. 
	 This scenario has been explicitly considered in both \cite{T15} and \cite{RB15}, in the latter of which an iterative method limited to adding two-particle sources was used, but in reverse: start from the network associated to the three-particle source and then connect one new observer via a two-particle source. However, we shall see that we can reproduce the quantum violations of both \cite{T15} and \cite{RB15} with our method.

	 The CHSH inequality\footnote{If an observer only has two measurement choices we will use the friendlier notations $a^k_0$ and $a^k_1$ instead of $a^k_\emptyset$ and $a^k_{\{1\}}$.} \cite{CHSH69} holds on $\mathscr{N}$;
	\begin{equation}
	\left\langle \frac{a^1_0+a^1_1}{2}a^2_0\right\rangle +\left\langle \frac{a^1_0-a^1_1}{2}a^2_1\right\rangle\leq 1. 
	\end{equation}
	We note that the connecting observer $A^2$ has $2^{s_2}$ with $s_2=1$ measurements, while we are adding $L=2$ observers. Therefore, using our theorem we find the following inequality for the network $\mathscr{N}'$:
	\begin{multline}\label{2p1}
	\min_{\sum q_X=1} \frac{1}{q_\emptyset} \left\langle \frac{a^1_0+a^1_1}{2}a^2_\emptyset \frac{b^1_0+b^1_1}{2}\frac{b^2_0+b^2_1}{2}\right\rangle \\
	+ \frac{1}{q_{\{1,2\}}}\left\langle\frac{a^1_0+a^1_1}{2}a^2_{\{1,2\}} \frac{b^1_0-b^1_1}{2}\frac{b^2_0-b^2_1}{2}\right\rangle\\ +\frac{1}{q_{\{1\}}}\left\langle\frac{a^1_0-a^1_1}{2}a^2_{\{1\}} \frac{b^1_0-b^1_1}{2}\frac{b^2_0+b^2_1}{2}\right\rangle\\
	+\frac{1}{q_{\{2\}}}\left\langle\frac{a^1_0-a^1_1}{2}a^2_{\{2\}} \frac{b^1_0+b^1_1}{2}\frac{b^2_0-b^2_1}{2}\right\rangle\leq 2.
	\end{multline}
	
	
	Let us investigate the possibility of violating the inequality using quantum theory. For this purpose, we introduce the states $|\phi_m\rangle=1/\sqrt{2}\left(|0\rangle^{\otimes m}+|1\rangle^{\otimes m}\right)$ and the mixture of $|\phi_m\rangle$ with random noise; $\rho_m=v|\phi_m\rangle\langle \phi_m|+(1-v)\textbf{1}/2^m$ for $v\in [0,1]$, where $\textbf{1}$ is the identity operator. We will denote the visibility in the network by $V$ which is the product of the $v$'s of all individual sources in $\mathscr{N}'$. We quantify the strength of a violation by the critical visibility $V_c$ which is defined as the largest value of $V$ that satisfies the inequality.
	
	For our example in Figure \ref{Examples} a), we distribute the state $\rho_2$ and $\rho_3$ respectively in the two sources in $\mathscr{N}'$. Let observers $A^1$, $B^1$ and $B^2$ perform measurements $a^1_0=b^1_0=b^2_0=\frac{\sigma_x+\sigma_y}{\sqrt{2}}\equiv M_+$ and  $a^1_1=b^1_1=b^2_1=\frac{\sigma_x-\sigma_y}{\sqrt{2}}\equiv M_-$, where $\sigma_x=|1\rangle\langle 0|+|0\rangle\langle 1|$ and $\sigma_y=i|1\rangle\langle 0|-i|0\rangle\langle 1|$. Let $A^2$ perform measurements $a^2_\emptyset=-a^2_{\{1,2\}}=\sigma_x\otimes\sigma_x$ and $a^2_{\{1\}}=a^2_{\{2\}}=\sigma_y\otimes \sigma_y$. A straightforward calculation will show that each of the four correlators in \eqref{2p1} takes the value $V/2\sqrt{2}$. Therefore, it is easy to realize that the minimum over $\sum q_X\!=\!1$ occurs at $q_X\!=\!1/4$ $\forall X\subset \{1,2\}$. Thus, we have found the violation $4\sqrt{2}V\nleq 2$ whenever $V>1/2\sqrt{2}$. The critical visibility $V_c=1/2\sqrt{2}$ is the same as obtained in previous studies of correlations on this network \cite{T15, RB15}. 
	
	\textit{Example 2.---}	
	Let us now consider the family of networks, here denoted $\mathscr{N}^{(N)}$, studied in \cite{T15} in which a center node is connected via $N$ sources to $L$ observers respectively, see Figure \ref{Examples} b). The intial network $\mathscr{N}^{(1)}$ is the standard Bell experiment in which observers $A^{1,1},...,A^{1,L+1}$ share a physical system. We can use our theorem $N-1$ times, each time adding a source connecting $A^{1,L+1}$ to $L$ new observers, until we have constructed $\mathscr{N}^{(N)}$.  
	
	We begin the iteration from the Bell inequality given in \cite{T15} for $\mathscr{N}^{(1)}$:	
 \begin{equation}
 \sum_{X\subset \mathbb{N}_L} \left\langle \prod_{k=1}^{L}\frac{a^{1,k}_0+(-1)^{\delta^k_X}a^{1,k}_1}{2}a^{1,L+1}_X\right\rangle\leq 1.
 \end{equation}

	Using $N-1$ iterations of our theorem to in every step add another $L+1$-particle source, we find that for some non-negative values of $\{q^l_X\}_X$ for $l=1,...,N-1$ such that $\sum q_X^l=1$ $\forall l$, there is a Bell-type inequality: 
 \begin{equation}\label{ineq2}
 \sum_{X\subset \mathbb{N}_L} \frac{1}{q^1_X\ldots q^{N-1}_X}\left\langle \prod_{j=1}^{N}\prod_{k=1}^{L}\frac{a^{j,k}_0+(-1)^{\delta^k_X}a^{j,k}_1}{2}a^{1,L+1}_X\right\rangle\leq 1.
 \end{equation}
	The correlators appearing in this inequality are of the same type as those that appears in the work of Ref.\cite{T15}. Following the analysis of Ref.\cite{T15}, we distribute the state $\rho_{L+1}$ in all $N$ sources. The observers $A^{j,k}$ for $j=1,...,N$ and $k=1,...,L$ perform $a^{j,k}_0=M_+$ and $a^{j,k}_1=M_-$. For the center node $A^{1,L+1}$; if $|X|$ is even (odd) then perform $a^{1,L+1}_X=(-1)^{\text{sg}_e(X)}\sigma_x^{\otimes N}$ ($a^{1,L+1}_X=(-1)^{\text{sg}_o(X)}\sigma_y^{\otimes N}$) where we have defined $\text{sg}_e(X)=(|X|\mod{4})/2$ and $\text{sg}_o(X)=((|X|-1)\mod{4})/2$. The work of \cite{T15} showed that such measurements result in each of the correlators in \eqref{ineq2} being $V/\sqrt{2^{nL}}$. The minimum over each of the sets $\{q^l_X\}_X$ for $l=1,...,N-1$ is achieved at $q^l_X=1/2^L$ for all $X$ and $l$. Thus, we find a violation $\sqrt{2^{nL}}V\nleq 1$ whenever $V>1/\sqrt{2^{nL}}$. The critical visibility $V_c=1/\sqrt{2^{nL}}$ is the same as found in Ref.\cite{T15} for this family of networks.

	\textit{Example 3.---}
	Let us now give an example going beyond the scope of previous work. Our network $\mathscr{N}''$ is given in Figure \ref{Examples} c). As initial network $\mathscr{N}$ we use the three-particle Bell experiment involving observers $A^1,A^2,A^3$. In a first use of our theorem, we connect $A^3$ to $B^1$ via a two-particle source,  forming the network $\mathscr{N}'$. Then, we connect $B^1$ via a three-particle source to observers $C^1$ and $C^2$, obtaining the network $\mathscr{N}''$. Since the network $\mathscr{N}'$ is the same as the network in Figure \ref{Examples} a), we can apply our theorem to the inequality \eqref{2p1} to obtain a Bell-type inequality for $\mathscr{N}''$.  
	This can compactly be written as 
	\begin{multline}\label{2p2p2}
	\min_{\sum q_X=1}\min_{\sum p_Y=1}\sum_{X\in \{1,2\}}\sum_{Y\in\{1,2\}}\frac{(-1)^{|X||Y|}}{q_Xp_Y}  \\
	\!\!\!\left\langle \prod_{k=1}^{2}\frac{a^k_0+(-1)^{\delta^k_X}a^k_1}{2} a^3_X b^1_Y \prod_{l=1}^{2}\frac{c^l_0+(-1)^{\delta^l_Y}c^l_1}{2}\right\rangle \leq 8
	\end{multline}

	To violate the inequality with quantum theory, we distribute the states $\rho_{3},\rho_2,\rho_3$ in the respective sources. We let observers $A^1,A^2,C^1$ perform the measurements $M_{\pm}$ (recall the definition of $M_{\pm}$ from \textit{Example 1}), while $C^2$ performs $c^2_0=\sigma_x$ and $c^2_1=-\sigma_y$. The observer $A^3$ performs $a^3_\emptyset=-a^3_{\{1,2\}}=\sigma_x\otimes \sigma_x$ and $a^3_{\{1\}}=a^3_{\{2\}}=\sigma_y\otimes \sigma_y$, and observer $B^1$ performs $b^1_{\emptyset}=-b^1_{\{1,2\}}=M_+\otimes M_+$ and $b^1_{\{1\}}=b^1_{\{2\}}=M_-\otimes M_-$. Calculating all the 16 correlators appearing in \eqref{2p2p2}, we find that for given $X,Y$, the associated correlator is $(-1)^{|X||Y|}V/4\sqrt{2}$. Thus, all the signs in the left-hand side of \eqref{2p2p2} cancel, making the minimization problem trivial; $q_X\!=\!p_Y\!=\!1/4$ $\forall X,Y$, leading to the violation $32\sqrt{2}V\nleq 8$ whenever $V>1/4\sqrt{2}$. The critical noise $V_c=1/4\sqrt{2}$ follows the Mermin-type scaling encountered for quantum correlations obtained from dichotomic measurements \cite{M90, BRGP12, TSCA14, RB15, T15}; that every additional observer amounts to an additional factor of $1/\sqrt{2}$ in $V_c$.

	\textit{Example 4.---} 
	As a final example, let us consider the network $\mathscr{N}''$ in Figure \ref{Examples} d). For the initial network $\mathscr{N}$ involving only the observers $A^1,A^2,A^3$, the Mermin inequality holds;
		\begin{equation}\label{Mermin}
		\left\langle \frac{a^1_0a^2_1+a^1_1a^2_0}{2}a_0^3\right\rangle +\left\langle \frac{a^1_0a^2_0-a^1_1a^2_1}{2}a_1^3\right\rangle \leq 1.
		\end{equation}
	
	By adding a three-particle source, connecting $A^3$ to observers $B^1$ and $B^2$, we obtain the network $\mathscr{N}'$. Using our theorem, we find the following Bell-type inequality on $\mathscr{N}'$.
	\begin{equation}\label{middle}
	\min_{\sum q_X=1}	\sum_{X\subset \{1,2\}} \frac{1}{q_X} \left\langle C_X a_X^3 \prod_{k=1}^{2}\frac{b^k_0+(-1)^{\delta^k_X}b^k_1}{2}\right \rangle\leq 2
	\end{equation} 
	where we have for simplicity introduced $C_\emptyset=C_{\{1,2\}}=(a^1_0a^2_1+a^1_1a^2_0)/2$ and  $C_{\{1\}}=C_{\{2\}}=(a^1_0a^2_0-a^1_1a^2_1)/2$.
	
	We connect observer $B^2$ to two new observers $C^1$ and $C^2$ via another three-particle source and thus obtain the network $\mathscr{N}''$. Again, applying our theorem starting from the inequality \eqref{middle} we find Bell-type inequality on $\mathscr{N}''$:
	\begin{multline}\label{ineq4}
	\min_{\sum q_X=1} \min_{\sum p_Y=1}	\sum_{X\subset \{1,2\}}\sum_{Y\subset \{1,2\}} \frac{(-1)^{\delta^2_X|Y|}}{q_Xp_Y} \\
	\!\!\!\left\langle C_X a_X^3 \frac{b^1_0+(-1)^{\delta^1_X}b^1_1}{2} b^2_Y \prod_{k=1}^{2}\frac{c^k_0+(-1)^{\delta^k_Y}c^k_1}{2} \right \rangle\leq 8.
	\end{multline}
	
	We can violate this inequality using quantum theory. Distribute in each source the state $\rho_3$ and let observers $A^1,A^2,B^1,C^1$ perform measurements $a^1_0=a^2_0=b^1_0=c^1_0=M_+$ and $a^1_1=a^2_1=b^1_1=c^1_1=M_-$. Let $A^3$ perform $a^3_\emptyset=-a^3_{\{1,2\}}=\sigma_x\otimes \sigma_x$ and $a^3_{\{1\}}=a^3_{\{2\}}=\sigma_y \otimes \sigma_y$, let $B^2$ perform $b^2_\emptyset=-b^2_{\{1,2\}}=M_+\otimes M_+$ and $b^2_{\{1\}}=b^2_{\{2\}}=M_-\otimes M_-$, and let $C^2$ perform $c^2_0=\sigma_x$ and $c^2_1=-\sigma_y$. Then it is straightforward to find that the correlator in \eqref{ineq4} associated to a given pair $X,Y$ is $(-1)^{\delta^2_X|Y|}V/4$. Thus, we find the violation $64V\nleq 8$ whenever $V>1/8$. The critical visibility $V_c=1/8$ corresponds to a critical visibility of $v=1/2$ per source, which falls in line with the expected scaling of $V_c$.
	
	\textit{Discussion.---} We constructed an iterative method for deriving Bell-type inequalities on any network of observers and sources that does not contain cycles. We examplified the capacity of our method for several networks by studying violations obtained from quantum theory, from which we could both successfully reproduce known results and certify strong quantum correlations in previously unstudied networks. 
	
	An interesting issue is the matter of critical visibilities for quantum correlations. It is known that in Mermin's multipartite Bell inequalities \cite{M90}, the critical visibility scales with a factor of $1/\sqrt{2}$ for every additional observer. To the author's knowledge, other and more general multipartite Bell inequalities with two-outcome measurements such as WWZB-inequalities do not offer better noise tolerance for their maximal violations \cite{WW01, ZB02}. An analog observation of critical visibilities has been made also for quantum correlations on networks \cite{BGP10, BRGP12, TSCA14, RB15, T15}. So far, no example has been found in which our presented method can provide a lower critical visibility than what is expected from such scaling. Nevertheless, there is evidence indicating that this is not a fundamental property and that quantum correlations on networks ought to reveal additional interesting features compared to standard Bell experiments \cite{CASA11}. For further investigations, it would be of great interest to construct Bell-type inequalities on networks with more than two outcomes for all observers. This has indeed been called for in several previous works.  
	
	As a final remark, we note that our technique does not work for a network that features a cycle. To the best of the author's knowledge, no examples of such Bell-type inequalities are known. Finding such an inequality even for a simple network with three observers and three two-particle sources, one connecting every pair of observers, would be very interesting.   
	
	\textbf{Acknowledgements.---}
	The author thanks Antonio Ac\'in and Umesh Vazirani for feedback and discussions.

\end{document}